# Coherent Control of Light Scattering

Alex Krasnok[1*], and Andrea Alú[1*]

[1]Advanced Science Research Center, City University of New York, New York, NY 10031, USA

E-mail: akrasnok@gc.cuny.edu, aalu@gc.cuny.edu

Scattering of electromagnetic waves, i.e., changing the direction of wave propagation and power flow caused by elastic (preservation of frequency) interactions with an object or scatterer, lies at the heart of the most experimental techniques over the entire optical spectrum. The first experimental investigations of electromagnetic wave scattering were performed in 1886-1889 years by Heinrich Hertz in his famous experiments that proved the theory of electromagnetic waves established by James Clerk Maxwell 20 years earlier [1]. The theory and experiments have convincingly proven that light is an electromagnetic wave and since then our research in the field of light phenomena has not ceased. Another example of light scattering that we encounter in nature is the reddening of the Sun at sunset and the blue color of the daytime sky, explained by Lord Rayleigh in 1871, well before the rigorous theory of such scattering was published in 1908 by Gustav Mie [2].

In everyday life, we meet with simple scattering processes, such as specular reflection, absorption, diffuse scattering, focusing, and lensing. However, progress in nanofabrication and material science made it feasible to crafty design scatterers demonstrating unusual scattering effects, including non-radiating states[3], cloaking[4], [5], bound states in the continuum[6], exceptional points[7]. Understanding these scattering effects plays a fundamental role in modern optical physics and is vital for many applications relying on enhanced light-matter interactions, e.g., energy harvesting [8], sensing [9] and optical data processing [10], to mention just a few.

These examples show that the design of single scatterers can lead to the appearance of unique scattering properties and exotic light-matter interaction effects. About fifteen years ago, it has been proven that combining subwavelength scatterers into an architected material, i.e., a *metamaterial*, may allow the realization of electromagnetic properties much different than those of the constituent materials [11]–[20]. Such artificial materials provide exotic electromagnetic responses leveraging on the unusual scattering properties of collections of plasmonic or high-index nanoparticles to enhance light-matter interactions at subwavelength scales and gain control over linear and nonlinear fields [21]–[28]. 2D planarized metamaterials, *metasurfaces*, have allowed obtaining specific optical properties on demand by controlling the local field distribution on a thin surface for various applications, including most notably wavefront engineering [12], [26], [29]–[37].

It has been shown recently that another fundamental property of light – *coherence* – can provide one more degree of freedom in scattering engineering and control. Nowadays this research area is a subject of significant interest. The examples of coherent scattering effects include *coherent perfect absorption* (CPA), *virtual perfect absorption* (VPA), *CPA lasing*, *coherently enhanced wireless power transfer* (CWPT). Here, we discuss these coherent scattering phenomena in photonics. We first establish a unified description of such exotic scattering effects in electromagnetics and show that the origin of all these effects can be traced back to singularities of the scattering matrix. Then we discuss these effects paying particular attention to their physical essence and coherent nature.

**Poles and zeros of the $\hat{S}$ matrix**

To begin with, here we derive a formulation for the general problem of light scattering. It consists in finding a solution of Maxwell's equations describing spatial and temporal evolution of total (impinging and scattered) electric $\mathbf{E}(\mathbf{r},t)$ and magnetic $\mathbf{H}(\mathbf{r},t)$ fields in one, two, or three dimensions

$$\nabla \times \mathbf{E}(\mathbf{r},t) = -\dot{\mathbf{B}}(\mathbf{r},t), \quad \nabla \cdot \mathbf{D}(\mathbf{r},t) = \rho(\mathbf{r},t), \tag{1a}$$

$$\nabla \times \mathbf{H}(\mathbf{r},t) = \dot{\mathbf{D}}(\mathbf{r},t) + \mathbf{j}(\mathbf{r},t), \quad \nabla \cdot \mathbf{B}(\mathbf{r},t) = 0, \tag{1b}$$

with corresponding boundary conditions. Here $\mathbf{D} = \varepsilon_0 \mathbf{E} + \mathbf{P} = \hat{\varepsilon}\mathbf{E}$, $\mathbf{B} = \mu_0(\mathbf{H}+\mathbf{M}) = \hat{\mu}\mathbf{H}$, $\rho$ and $\mathbf{j}$ are the free charge density and current, $c$ is the speed of light in vacuum, $\varepsilon_0$ is the dielectric constant, $\hat{\varepsilon}$ and $\hat{\mu}$ are permittivity and permeability (tensors and/or integral operators, in general). The general boundary conditions can be written as $\mathbf{n} \times (\mathbf{E}_2 - \mathbf{E}_1) = 0$, $\mathbf{n} \cdot (\mathbf{D}_2 - \mathbf{D}_1) = \rho$, $\mathbf{n} \times (\mathbf{H}_2 - \mathbf{H}_1) = \mathbf{J}$, and $\mathbf{n} \cdot (\mathbf{B}_2 - \mathbf{B}_1) = 0$, with $\mathbf{n}$ being the normal vector from medium "1" to medium "2".

Hereinafter, we consider only uncharged, linear, nonmagnetic, and isotropic materials, when $\hat{\varepsilon}$ is reduced to one complex valued permittivity $\varepsilon(\omega)$. We also assume that the host medium permittivity $\varepsilon_h$ is a real valued constant. Moreover, we imply the time convention

$\exp(-i\omega t)$. Under these assumptions we derive the Helmholtz equation from Eqs. (1) in time-domain

$$\nabla \times \nabla \times \mathbf{E}(\omega,\mathbf{r}) - \varepsilon(\omega)\frac{\omega^2}{c^2}\mathbf{E}(\omega,\mathbf{r}) = 0. \quad (2)$$

The total electric field can be divided into the incident $\mathbf{E}_i$ and scattered $\mathbf{E}_s$ components, $\mathbf{E} = \mathbf{E}_i + \mathbf{E}_s$. Hence, the solution of Eq.(2) can be formally found in form $\mathbf{E}_s = \hat{S}\mathbf{E}_i$, with $\hat{S}$ being the *scattering matrix* (in operator form)[38].

Let's discuss the structure and the main properties of the $\hat{S}$ matrix. For this, we first chose a basis of certain incoming and outgoing waves, or *channels*, Fig. 1(a), and represent the incident and scattered waves in the form $\mathbf{E}_i = \sum_m s_m^+ \mathbf{E}_m^+$, and $\mathbf{E}_s = \sum_m s_m^- \mathbf{E}_m^-$, with corresponding amplitudes $\mathbf{s}^+ = \{s_1^+, s_2^+, ...\}$ and $\mathbf{s}^- = \{s_1^-, s_2^-, ...\}$, respectively. Hence, the formal solution to Eq. (2) in the basis of chosen channels can be presented in form

$$\mathbf{s}^- = \hat{S}\mathbf{s}^+, \quad (3)$$

where $\hat{S}$ is the scattering matrix of the system in a chosen basis. Thus, amplitudes of the outgoing waves $\mathbf{s}^-$ are linear combinations of the incoming wave amplitudes $\mathbf{s}^+$, which is established by the $\hat{S}$ matrix of the system. Usually, the amplitudes are normalized such that $|s_m^+|^2$ and $|s_m^-|^2$ correspond to the energy of the incoming and outgoing waves in a certain channel $m$. For example, in a single port system, the $\hat{S}$ matrix coincides with the reflection coefficient ($r$). In a two-port system, the scattering matrix is $\hat{S} = \begin{pmatrix} r_{11} & t_{12} \\ t_{21} & r_{22} \end{pmatrix}$, where $r_{ii}$ and $t_{ij}$ stand for corresponding reflection and transmission coefficients. The channels exist outside the system and represent freely propagating solutions of the wave equation in the absence of the scattering object. The specific choice of channels often depends on the exact geometry of the system. For instance, in a planar slab the basis of plane waves is the obvious choice, whereas for a spherical particle it is convenient to use a basis of vector spherical harmonics.

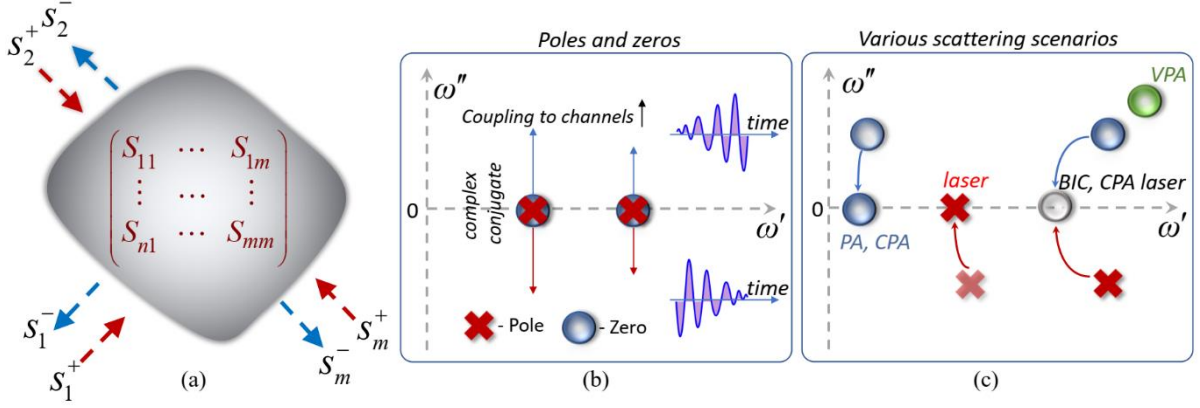

**Figure 1. Scattering matrix and its zeros and poles.** (a) Description of a linear electromagnetic structure by a scattering ($\hat{S}$) matrix, which links the amplitudes of the ingoing ($s_m^+$) and outgoing ($s_m^-$) waves; $m$ stands for the number of channels. (b), (c) Various cases of light scattering scenario and their relation to poles and zeros location. (b) Hermitian system. The zeros and poles are related through complex conjugation. In the limit case of almost no coupling to channels, pole and zero lie very close to the real axis. With increase in coupling to channels (radiative losses), the pole (zero) goes to upper (lower) complex plane. (c) Non-Hermitian scenarios, including lasing, perfect absorption (PA), coherent perfect absorption (CPA). BIC stands for bound state in the continuum, the situation when the zero and pole of a Hermitian system coalesce at the real axis. VPA stands for virtual perfect absorption. CPA-laser regime is possible in a system with balanced loss and gain.

Now we discuss the general properties of scattering matrix. If a system is *reciprocal* [39]–[44], it implies that the scattering matrix is symmetric, $\hat{S} = \hat{S}^T$. Next, usually the $\hat{S}$ matrix can be reduced to its diagonal form $\hat{S} = \text{diag}(d_1, d_2, \cdots, d_m)$, where each $\hat{S}$ matrix *eigenvalue* $d_p(\omega)$ is associated with its incoming amplitudes (eigenvectors) $\mathbf{s}_p^+$, $\hat{S}\mathbf{s}_p^+ = d_p \mathbf{s}_p^+$. In a *Hermitian system*, i.e., a system with no gain and material loss (the permittivity is real everywhere), the scattering eigenvalues $d_p$ *at real frequency* $\omega'$ are unimodular, i.e., $|d_p(\omega')| = 1$. This constraint is imposed by the unitarity of the scattering matrix, i.e., $\hat{S}\hat{S}^+ = \hat{I}$, which is associated with a Hermitian Hamiltonian $\hat{H}_0 = \hat{H}_0^+$ of the scatterer[45], [46]. The eigenvalues of $\hat{S}$ matrix of a Hermitian system can be presented in the exponential form $d_p(\omega') = \exp[2\iota\delta_p(\omega')]$ which shows that a corresponding eigenvector $s_p^+$ experiences a phase shift $\delta_p(\omega')$ upon scattering on the system.

When the frequency $\omega$ is analytically continued to the complex plane, $\omega = \omega' + i\omega''$, the scattering eigenvalues $|d_p(\omega)|$ may take any values between $\infty$ or 0. The values $\infty$ and 0 correspond to the poles and zeros of the $\hat{S}$ matrix respectively. The poles relate to the *eigenmodes* (self-suspended solutions)[47] while the latter is associated with the *perfect absorbing states* (no scattering). A fundamental property of the $\hat{S}$ matrix eigenvalues $d_p$ is that they can be expanded into a special product via the *Weierstrass factorization theorem* [48]–[50]:

$$d_p(\omega) = A_p \exp(iB_p\omega) \prod_m \frac{\omega - \omega_{zr,m}}{\omega - \omega_{pl,m}}, \qquad (5)$$

where $A_p$ and $B_p$ are constants, and the product is taken over all resonances which match the symmetry of $d_p$. As one can directly see, these expressions include only information about positions of poles $\omega_{pl,m}$ and zeros $\omega_{zr,m}$ in the complex plane. Therefore, the Weierstrass expansion shows that the scattering process is *fully described only by the positions of poles and zeros*. This fact makes it possible to describe and classify a variety of scattering phenomena in terms of the position of zeros and poles.

In lossless systems, poles $\omega_{pl,n}$ and zeros $\omega_{zr,n}$ occur in *complex conjugate pairs*, i.e., $\omega_{zr,n} = \omega_{pl,n}^*$ and restricted to specific parts of the complex frequency plane. Resonant states of a passive system must decay in time due to leakage of the energy from the system into free space. With the time dependence defined through the factor $\exp(-i\omega t)$, such attenuation corresponds to the complex frequencies $\omega_{pl,n}$ in the lower half-plane. Correspondingly, the zeros $\omega_{zr,n} = \omega_{pl,n}^*$ sit symmetrically in the upper half-plane, Fig. 1(b). In the limit of zero coupling, poles coincide with the eigenfrequencies of the close-system Hamiltonian, $\hat{H}_0$. When the coupling between the scatterer and the channels increases, the pairs of pole and zero start to move apart from the real axis of the complex frequency plane, Fig. 1(b). Note that the coalescence of a pole and zero can happen not only in a closed system but also in so-called *bound state in the continuum* (BIC), Fig. 1(c), which is an eigenmode of the scatterer with a real frequency in the continuum of unbounded modes [6], [51]–[53].

If a system is non-Hermitian ($\hat{H}_0 \neq \hat{H}_0^+$) the poles and zeros are located asymmetrically in the complex frequency plane and their position can be controlled by material gain and loss parameters, which *break the time-reversal symmetry* $\mathcal{T}$. In pure lossy scatterers ($|d_m(\omega)|^2<1$, for each channel), the complex zeros move down from the upper-frequency plane as dissipation increases. At some critical value of dissipation, the scattering zero crosses the real axis at some frequency leading to *perfect absorption* (PA) of an appropriate incoming waveform, Fig. 1(c). This waveform is determined by the eigenstate of $\hat{S}(\omega)$ corresponding to its zero eigenvalue. In the case of several ports, it gives rise to the coherent perfect absorption (CPA), the time-reversed counterpart of a laser for a structure with gain[54], [55]. If the system has only one port, this condition corresponds to a common perfect absorption regime.

For a given system with some amount of losses, the increase in gain allows compensation of the energy loss. In the complex plane, it corresponds to pushing a pole closer to the real axis, Fig. 1(c). As the material part of loss compensated, the further growth of gain allows *amplification regime* and a signal passing through the system gets amplified. The *lasing threshold* can be reached when further gain enhancement compensates the entire losses (material and radiation) in the system. This is accompanied by the phase transition to a coherent light emission[56]. Thus, in the complex plane, lasing corresponds to the presence of a pole at the real axis[47], Fig. 1(c). It worth to mention that although the *laser generation is essentially a nonlinear process*, the threshold pumping for lasing can be found within a linear approximation[47], [57] because the amplitude of the laser self-oscillation is equal to zero below the lasing threshold and the problem can be treated linearly.

The balance of gain and loss leads to the $\mathcal{PT}$-symmetric system, which evolution is governed by a Hamiltonian symmetrical upon reversal of space $\mathcal{P}$ and time $\mathcal{T}$. The PT-symmetric optical system has balanced absorption and amplification as long as the relative permittivity of the medium satisfies $\varepsilon(\mathbf{r})=\varepsilon^*(-\mathbf{r})$. Such $\mathcal{PT}$-symmetric Hamiltonians support the phase transition of eigenvalue spectrum from entirely real to complex ones, which occur as the so-called *exceptional points* (EPs)[58]–[63]. EPs emerge when two (or more) eigenvalues and their corresponding eigenstates coalesce simultaneously so that the Hamiltonian becomes defective. In an the interesting subclass associated with wave scattering due to $\mathcal{PT}$-symmetric targets, like the Hermitian case, poles and zeros also occur in complex conjugate pairs, as seen from the interchange between the secular equation for the poles and zeros after performing the combined $\mathcal{PT}$ symmetry operation. Nevertheless, as the strength of the non-Hermiticity varies, pairs of pole and zero will move symmetrically in the complex frequency plane, Fig. 1(c). Eventually, a pole and a zero of the $\hat{S}$ matrix may coincide on the real axis and a *CPA-laser*[64]–[66] point emerge, when the system behaves simultaneously as a laser oscillator and a CPA.

Note that, besides the poles and zeros of the $\hat{S}$ matrix, one can use poles and zeros of other physical quantities for analysis and designing of scattering at will. For example, in the 1D case, the reflection coefficient also has these singularities in the complex frequency plane. Interestingly, poles of the scattering amplitude coincide with those that of $\hat{S}$ matrix eigenvalues. However, in general, zeros do not coincide with zeros of $\hat{S}$ matrix since they depend on excitation geometry (angle of incidence, number of excitation beams). This gives rise to the *coherent control of scattering* when one uses two or more coherent light beams in a two or more channel configuration.

Another example of scattering poles and zeros we meet in 2D and 3D case. For analysis of such structures, we often use the basis of spherical or cylindrical channels (harmonics) with the corresponding scattering coefficients. Let's consider here the case of the 3D object of spherical symmetry of a radius $R$. In this case, the Mie scattering theory can be used, which yields the following equation for the scattering cross section (SCS)[67], [68]

$$SCS = \frac{\lambda^2}{2\pi} \sum_{l=1} (2l+1)\left(|c_l^{\text{TM}}|^2 + |c_l^{\text{TE}}|^2\right), \tag{6}$$

with $c_l^{\text{TM}}$ and $c_l^{\text{TE}}$ being the scattering coefficients (Mie coefficients), which can be calculated analytically[67]. In Eq.(6) $l$ defines the order of a scattering channel and $\lambda$ is the wavelength of impinging light. The scattering coefficients can be expressed as $c_l^{\text{TE,TM}} = -U_l^{\text{TE,TM}} / (U_l^{\text{TE,TM}} - jV_l^{\text{TE,TM}})$ [69], for TE and TM polarized waves, respectively. The quantities $U_l^{\text{TE,TM}}$ and $V_l^{\text{TE,TM}}$ are defined by a combination of spherical Bessel and Neumann functions[70]. The scattering coefficients indicate the contribution of the different scattering channels, and the superscript TM or TE indicates whether the vector spherical harmonic have magnetic or electric field orthogonal to the radial direction.

If a scatterer is subwavelength ($kR \leq 1$), we may cut Eq. (6) at the first dipole term ($l = 1$). In this case, the only one dipole scattering amplitude $c_1^i = -U_1^i / (U_1^i - jV_1^i)$ governs the scattering properties. This amplitude has a *scattering zero* when $U_1 = 0$ giving rise to the *anapole state*[3], which is relative to the cloaking effect[5] for a subwavelength scatterer. The anapole state is of great interest in modern photonics and has been employed for boosting nonlinear effects in dielectric nanoparticles, such as third harmonic generation and four-wave

mixing[71]–[74]. In its turn, the *scattering poles* (eigenmodes) are realized when the denominator turns to zero, $U_1^i - jV_1^i = 0$. In such a structure, adding material gain can also push the pole to the real frequency axis leading to realization of lasing regime (nanolasers, spacers)[75]. The pole and zero of the 3D structure can also coalesce at the real axis by tailoring the structure geometry with realization of the bound state in the continuum (BIC) scattering regime[52].

**Coherent perfect absorption (CPA)**

Scattering of any linear structure is determined by two basic kinds of peculiarities in the complex frequency plane, zeros and poles of the $\hat{S}$ matrix eigenvalues located in the upper and lower half-planes, respectively, see for example Fig. 2(a). These singularities corresponding to perfectly absorbing and lasing solutions can never be achieved by *monochromatic excitation* in a Hermitian system. However, the presence of material losses inevitably breaks the time-reversal symmetry $\mathcal{T}$ and a specific amount of material loss can get one of the zeros to the real frequency axis, Fig. 2(b). In this case, the system would work as a perfect absorber (PA) for excitation by monochromatic waves in an *eigenvector configuration*. In some sense, this regime can be considered as time-reversed lasing, when adding gain to the system pushes poles towards the real axis until one of them get to the real axis with formation of self-sustained oscillations of the electromagnetic field without incident field at the same frequency[47], [76].

The vast majority of perfectly absorbing systems utilizes the one port excitation geometry when the incident energy is delivered to the system through a single channel. This situation corresponds to the most practical case of an absorber illuminated from one side by, e.g., the Sun, radar or a lab light source. For example, classical one-port perfect absorbers, *Salisbury* and *Dallenbach screens* [77], [78] utilize the destructive self-interference of a plane wave excitation on one side of an absorber. More recent examples include PAs based on diffraction gratings and metasurfaces[79]–[88]. In order to enable destructive interference, an absorbing material should be placed above a mirror or covered with an anti-reflective coating[89]. Thus, achieving one-sided perfect absorption without backing mirrors is a challenging task, especially in the visible range.

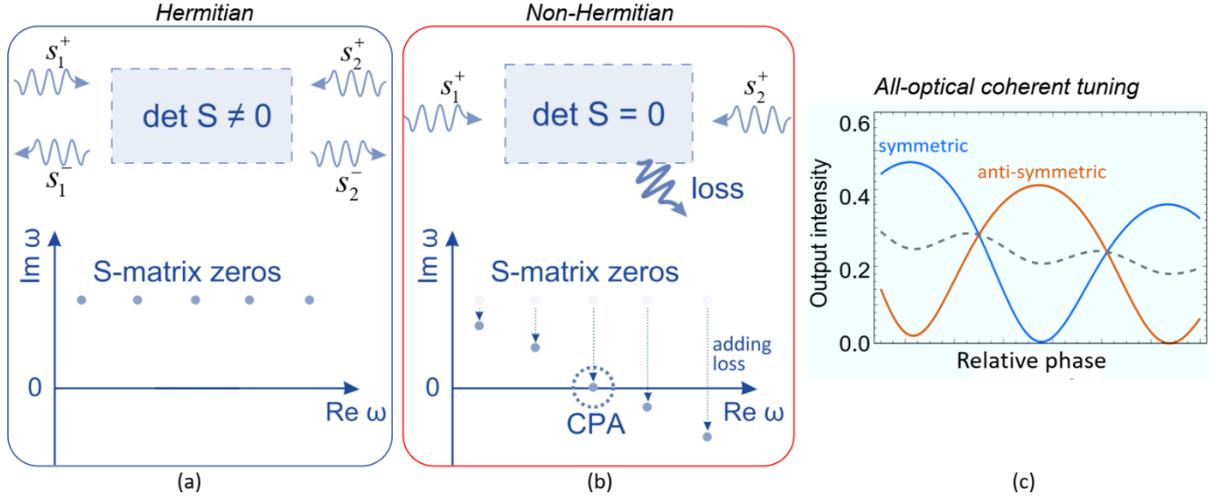

**Figure 2. Coherent perfect absorbers.** (a) Hermitian two-port system (e.g., Fabry-Perot resonator). The $\hat{S}$ matrix zeros of such a system lie in the upper complex plane. Adding material loss to the system pushes the zeros down towards the real axis (b). As one of them get to the real axis, the system becomes CPA at the corresponding frequency. (c) Tuning of phase or amplitude of one of the port signals allows tuning the output intensity of the system.

In a system with two or more excitation channels, the perfect absorption concept is generalized to coherent perfect absorption (CPA). In this scenario, if the system is excited through two or more channels with appropriate amplitudes and phases (eigenvectors of $\hat{S}$ matrix), it perfectly absorbs all energy [54], [55], [90]–[92]. This phenomenon can be illustrated by consideration of a two-port planar structure, Fig. 2. The inputs and outputs of such system are related via the $\hat{S}$ matrix (for normal incidence), $\begin{pmatrix} s_1^- \\ s_2^- \end{pmatrix} = \hat{S} \begin{pmatrix} s_1^+ \\ s_2^+ \end{pmatrix}$, where $\hat{S} = \begin{pmatrix} r_{11} & t_{12} \\ t_{21} & r_{22} \end{pmatrix}$. Here $s_i^+$ and $s_i^-$ respectively denote the input and output wave amplitudes in the i-th channel; $r_{ii}$ are the reflection coefficients in each port, and $t_{ij}$ are the transmission coefficients. This $\hat{S}$ matrix has the following eigenvalues and eigenvectors

$$d_{1,2} = \frac{1}{2}\left(r_{11} + r_{22} \pm \sqrt{r_{11}^2 - 2r_{11}r_{22} + r_{22}^2 + 4t_{12}t_{21}}\right), \tag{7a}$$

$$\mathbf{s}_{1,2} = \left\{1, -\frac{-r_{11} + r_{22} \pm \sqrt{r_{11}^2 - 2r_{11}r_{22} + r_{22}^2 + 4t_{12}t_{21}}}{2t_{21}}\right\}, \tag{7b}$$

If $\hat{S}$-matrix is constrained by optical reciprocity[41], then $t_{12} = t_{21} = t$. If in addition a two-port structure is symmetric under mirror reflection, $r_{11} = r_{22} = r$, the corresponding eigenvalues and eigenvectors become $d_{1,2} = r \pm t$ and $\mathbf{s}_{1,2} = \{1, \pm 1\}$ representing *symmetric and antisymmetric inputs* of equal intensity.

If the system is lossless, all zeros of $d_{1,2} = r \pm t$ are located in the upper complex plane, Fig. 2(a). However, adding a certain amount of material losses can make $r \pm t = 0$ at the real frequency axis, Fig. 2(b). Then illumination of it with the corresponding CPA eigenmode $\mathbf{s}_{1,2} = \{1, \pm 1\}$ will lead to complete absorption. The absorption in a CPA turns out to be highly sensitive to variations in the conditions of illumination, in particular in the relative phase between the components of the exciting field. To characterize a CPA it is convenient to define the *joint absorption* [93]

$$\mathcal{A} \equiv 1 - \frac{|s_1^-|^2 + |s_2^-|^2}{|s_1^+|^2 + |s_2^+|^2}. \tag{8}$$

Here, $|s_1^+|^2 + |s_2^+|^2$ is proportional to the total input intensity, and $|s_1^-|^2 + |s_2^-|^2$ to the total output intensity; thus a CPA has $\mathcal{A} = 1$. Fig. 2(c) features a typical plot of $1 - \mathcal{A}$ for the symmetric (blue curve) and anti-symmetric (orange curve) modes of a uniform dielectric slab near a CPA frequency. CPA occurs whenever one of the curves falls to zero. An intriguing feature of this plot is that at the CPA frequency, while one of the incident modes yields total absorption, the other mode strongly scatters due to constructive interference outside the system, thus suppressing the absorption. By adjusting the relative phase of the inputs, one can switch to the regime of suppressed absorption, caused by the constructive interference of exiting waves [54], [91], [93].

Zeros of different eigenvalues of $\hat{S}$ matrix typically occur at distinct frequencies, meaning that if a certain incident wave experiences CPA, the orthogonal wave does not. However, there are examples of doubly degenerate CPA[94], [95], when both eigenvalues of $\hat{S}$ can be zero at the same frequency. In this case the absorption becomes insensitive to the relative phase of the two input waves, so that even a single input wave incident on either port would be perfectly absorbed.

CPAs have been studied and realized in a wide variety of geometries, including planar structures (slabs, diffraction gratings, metasurfaces, and thin films) [54], [90], [91], [96] and guided-mode structures [97]–[99] compatible with integrated nanophotonics applications. Different techniques have been devised for increasing or decreasing the absorption band-width[100]–[102], and for achieving perfect absorption under strongly-asymmetric beam intensities[103]. Also CPA has been extended to systems exhibiting strong light-matter interaction[104]. Recently, CPA has been generalized to disordered systems[92].

The phenomenon of CPA may be useful in a variety of applications, ranging from all-optical data processing to enhanced photocurrent generation[55]. CPAs can be highly sensitive to the parameters of the input waves, enabling attractive opportunities for the flexible control of light scattering and absorption[91], [105], [106]. For instance, in a two-port CPA, one input can be regarded as a signal beam, and the other as a control beam. A coherent control beam serves as a resource enabling efficient control of optical absorption, even in the linear regime. This opens the door to novel low-energy logical circuits and small signal amplifiers[105], [106] without the use of non-linear materials that rely on high intensity of incident radiation to achieve non-linear response. Moreover, coherent perfect absorption can be employed for processing of binary images and pattern recognition[107]. Coherent absorption effects may prove very useful for improving photocurrent generation in photoelectrochemical systems[108]. Finally, the ability to control and remotely trigger coherent perfect absorption of single- or few-photon states[109], [110] could be promising for quantum information processing.

**Virtual perfect absorption (VPA)**

Absorbing electromagnetic waves is crucial to many problems of applied science, including sensing and molecular detection, RF cloaking, photovoltaics and photodetection. Crafty designed perfect absorbers (PA) allow light absorption in an ultimately effective fashion boosting the performance of devices relying on light absorption, e.g. ultrathin solar cells. CPA is a two or more port absorber with, as in the case of common PAs, zeros of $\hat{S}$ matrix eigenvalues at the real frequency axis. Although, CPAs allow perfect absorption and all-optical coherent light manipulation, the energy is eventually dissipated into heat, which is undesirable for some applications in active light manipulation such as low-energy memories and optical modulators. For these applications, it is important to design a system capable of storing the absorbed energy for an arbitrary amount of time and releasing it at will, rather than its dissipation. Conventional high-Q lossless cavities such as microdisks[111]–[115], microspheres[116], Bragg reflector microcavities[117], and photonic crystals[118]–[120] with extremely high Q factors ($\sim 10^3 - 10^6$) are capable of storing energy for a relatively long time however the process of charging these cavities are highly inefficient. In these system, at every cycle of the external excitation, huge amount of energy is reflected and only a tiny part of the illumination penetrates into the cavity, resulting in a very inefficient storing process. Although, to perfectly couple to an optical resonator the so-called *critical coupling* (often called as Q-factor matching) approach can be used, this require presence of losses (material or radiative) which is very undesirable in applications. In other words, to critically couple to a mode with a large $Q$-factor one needs to add losses into it and hence spoil the cavity.

Recently, the concept of coherent perfect absorption has been generalized to the *complex frequencies* and *Hermitian systems*. Instead of adding loss in order to push the complex zeros of the system towards the real frequency axis, it is feasible to engage the complex zeros of a lossless system by tailoring a signal with the corresponding complex frequency[121]. It was shown that if a lossless dielectric slab is coherently excited with signals with these complex frequencies [Fig. 3(a)], the energy of the incoming waves will get stored in the slab *without any outgoing energy from the system*, Fig. 3(b). Fig. 3(c) shows the stored energy during this temporal process. Fig. 3(d), (e) demonstrate the electric field distribution in the system during the excitation (d) and right after excitation breaking (e). Hence, during this excitation by a complex signal the structure looks like perfectly absorbing. As soon as the illuminating signals stop, all the energy stored in the system will get released through its electromagnetically open channels.

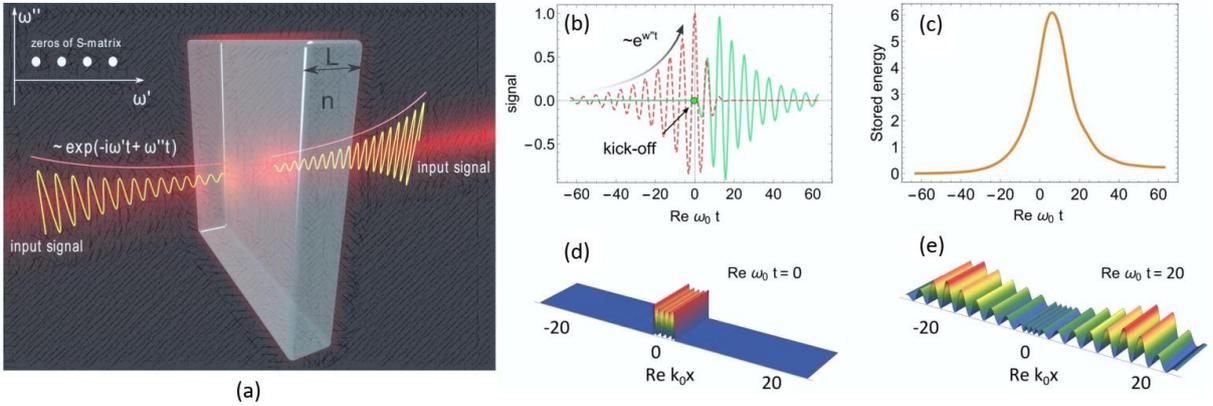

**Figure 3. Virtual perfect absorption.** (a) Dielectric slab resonator is coherently excited with signals corresponding to one of the zeros in the upper complex plane. (b) During the excitation by a signal corresponding to one of the zeros (red curve), the total scattering to all ports is absent (green curve). As the excitation stops abruptly, the stored energy gets released. (c) Stored energy during this temporal process. (d), (e) Electric field distribution in the system during the excitation (d) and right after excitation breaking (e).

Thus, instead of adding loss to the system to push the complex zero towards the real axis, it is possible to tailor the incident field in time, such that its temporal profile matches the exponentially diverging mode associated with the complex zero over a finite interval of time, giving rise to *virtual perfect absorption*[121]. The effect can be reproduced in any lossless (or with small loss) electromagnetic structure provided that the spectral position of its scattering zero is known and the corresponding spatial and temporal profile of the incident field can be created. It is also robust to inevitable material dissipation and material dispersion because the position of poles and zeros entails information about this dispersion.

**Coherently enhanced wireless power transfer (CWPT)**

An antenna is a crucial element for many important wireless technologies, including communications and power transfer[122]. First antennas emerged at the same time with the discovery of electromagnetic waves by H. Hertz in 1888 and since then have been developing with human civilization often being a catalyst for its development. A plethora of antennas have been invented for the radio, microwave, THz and even optical frequency range, where they become unique elements for quantum optics and interconnections on a chip[123]–[125].

While wireless communications are rather established, the wireless power transfer (WPT), proposed at the beginning of the 20th century by N. Tesla[126] is experiencing a rebirth. It was caused by an experimental demonstration that the WPT efficiency, i.e. amount of energy received by a reviving antenna related to the total amount of the emitted energy by transmitter (usually a magnetic coil), can be drastically enhanced in so-called near-field WPT regime, when the power is transferred via resonant coupling[127]. Transfer over the distance of 2 m with high efficiency in the kHz range via strongly coupled magnetic resonances between two metallic coils has been achieved[127] that gave rise to research on ways of using this effect for novel technologies. Examples include electric vehicles, implanted medical devices, and consumer electronics. Since then, significant research efforts have been devoted to exploring the ways to achieve as high WPT efficiency as possible[128], whereas the vast majority of them has been concentrated on optimizing of the resonators' geometry, a surrounding, and their relative arrangement.

In the previous subsections, it has been shown that electromagnetic absorption and scattering can be effectively controlled through tailoring of coherent temporal shaping of the incident electromagnetic field. Namely, a coherent perfect absorber (CPA) is a linear electromagnetic system in which perfect absorption is achieved with two or more incident waves, creating constructive interference inside an absorbing structure. Here we show that the same principle that underlies the operation of CPAs can be employed to improve the efficiency of antenna energy transfer, and as a result improve the efficiency of WPT systems[129]. More specific, by coherently exciting the receiving antenna [see Fig. 4(a)] with an auxiliary signal, tuned in sync with the impinging signal from the transmitting antenna, it is possible to enhance the robustness of the system largely. This additional signal improves energy transfer through constructive interference with the impinging wave, compensating any imbalance in the antenna coupling without having to modify the load.

The analysis of this system can be performed by the temporal coupled mode theory (TCMT)[130]–[132]. We assume that the dipole receiving antenna has a single mode with the real eigenfrequency $\omega_0$ and the mode amplitude $a$, normalized such that $|a|^2$ is the energy of the mode. The dipole antenna couples to the waveguide and free space radiation with the coupling constants $|\kappa\rangle = \{\kappa_w, \kappa_f\}$. The excited antenna mode can decay to both channels with the total dumping rate, $1/\tau = 1/\tau_w + 1/\tau_f$. In result we can write out the first equation of the TCMT describing the evolution of the dipole antenna mode amplitude in form

$$\frac{da}{dt} = \left(i\omega_0 - \frac{1}{\tau}\right)a + \langle \kappa^* | s_+ \rangle, \tag{9}$$

where $|s_+\rangle$ is the vector of input amplitudes. Now we assume that the vector of input amplitudes consists of not only the field of the transceiver $s_f$ but also the additional field of coherent excitation $s_w^+$, Fig. 4(a). As follows from Eq. (1), without excitation ($|s_+\rangle = 0$) the initially excited dipole mode decays in time as $a \sim e^{i\omega_0 t} e^{-t/\tau}$, i.e., exponentially with the complex frequency, $\omega = \omega' + i\omega''$, where $\omega' = \omega_0$ and $\omega'' = -1/\tau$.

In its turn, the output amplitude vector $|s_-\rangle$ is related to the input vector and the mode amplitude via

$$|s_-\rangle = \hat{C}|s_+\rangle + a|\kappa\rangle \tag{10}$$

where $\hat{C}$ is the matrix of *direct scattering*. We assume the waveguide is highly mismatched with the free space and hence $\hat{C} = -1$.

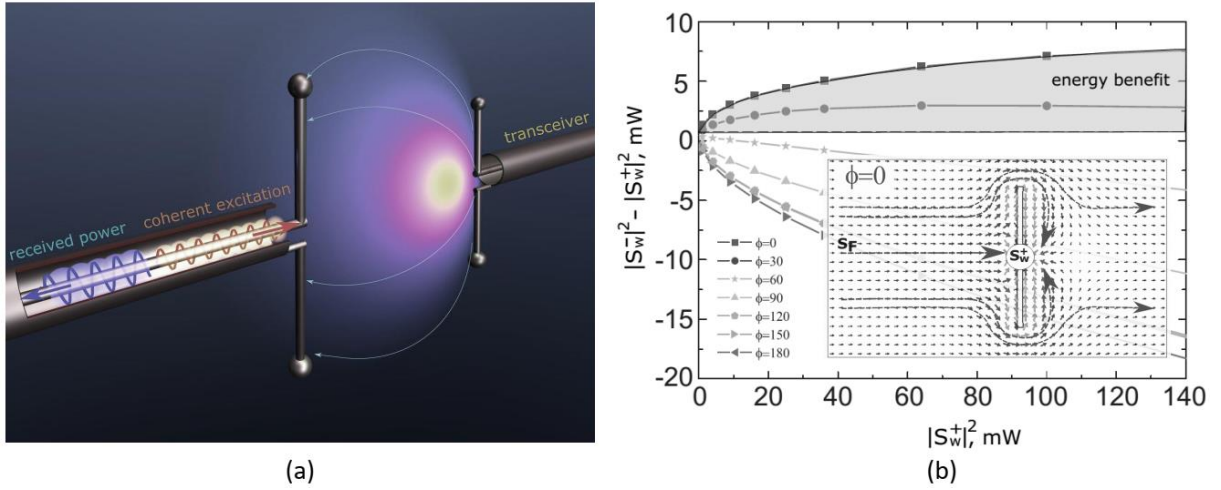

(a)                         (b)

**Figure 4. Coherently enhanced wireless power transfer.** (a) Schematic of a coherently enhanced wireless power transfer system consisting of a transceiver antenna and a receiver antenna driven by an additional coherent excitation. (b) Net extracted energy as a function of the auxiliary signal intensity for different relative phase values. Filled area indicates the region of energy benefit, where the coherently assisted energy balance exceeds that for $s_w^+ = 0$. Inset: Poynting vector distribution around the antenna with the auxiliary signal $s_w^+$ and relative phase of 0 deg.

The results of the TCMT analysis yields the following expression for the scattered energy back to the receiver (extracted field)[129]

$$s_w^- = -s_w^+ + \kappa_w \frac{\kappa_w s_w^+ + \kappa_f s_f}{i(\omega - \omega_0) + 1/\tau}. \tag{11}$$

If $s_f = 0$ (no radiation from the transceiver), this expression gives the reflection parameter ($s_{11}$), which can be measured or calculated. The typical results that this equation yields for the net extracted energy are shown in Fig. 4(b). The area where the net extracted energy exceeds that for $s_w^+ = 0$ (filled area) corresponds to the energy benefit. In these regimes, the antenna revives more energy from the transceiver than it does without coherent excitation even after subtraction of this additional energy. Interesting that in such a regime of positive net extracted energy the Poynting vector distribution around the antenna demonstrates many flow lines ending by the dipole antenna, Fig. 4(b, inset). Otherwise, when the relative phase between $s_w^+$ and $s_f$ is 180 deg., there are a few of Poynting vector lines flowing into the antenna that operates in the radiation regime (radiates more than receives). In a practical WPT device, the amplitude and phase of the additional coherent signal can be controlled in real time to adjust the antenna as a function of changes in the environment, temperature changes in the load, and distance of the transmitter.

## Conclusions

Scattering of electromagnetic waves lies at the heart of all experimental techniques in radiophysics, visible and X-ray optics. Hence, deep insight into the basics of scattering theory and understanding the peculiar features of electromagnetic scattering is necessary for the correct interpretation of experimental data and an understanding of the underlying physics. Recently, a broad spectrum of exceptional scattering effects, including bound states in the continuum, exceptional points in PT-symmetrical non-Hermitian systems, and many others attainable in wisely suitably engineered structures have been predicted and demonstrated. Among these scattering effects, those that rely on coherence properties of light are of a particular interest today. Here, we have discussed coherent scattering effects in photonics. Coherent perfect absorption (CPA) generalizes the concept of perfect absorption to systems with several excitation channels. This effect allows all-optical light manipulation by tuning of parameters of coherent excitation. In its turn, the virtual perfect absorption (VPA) effect generalizes the concept of CPA to complex frequencies and Hermitian systems. In contrast to real energy dissipation in PA and CPA, in VPA the energy is trapped in a resonator till excitation continuous exponentially growth and gets released when it breaks. Additionally, VPA provides a way to perfectly couple to any even lossless high-Q resonators. Finally, employing principles of coherent excitation can boost the performance of traditional WPT systems via coherently enhanced wireless power transfer (CWPT) effect.